\documentclass{egpubl}
\usepackage{eurovis2022}
\SpecialIssuePaper   
\usepackage[T1]{fontenc}
\usepackage{dfadobe}  
\usepackage{cite}  
\BibtexOrBiblatex  
\electronicVersion
\PrintedOrElectronic
\ifpdf \usepackage[pdftex]{graphicx} \pdfcompresslevel=9
\else \usepackage[dvips]{graphicx} \fi
\usepackage{egweblnk}
\usepackage{pdfpages}
\usepackage{chngcntr}
\usepackage{makecell}
\usepackage{hyperref}
\hypersetup{
    unicode=false,          
    pdftoolbar=true,        
    pdfmenubar=true,        
    pdffitwindow=false,     
    pdfstartview={FitH},    
    pdftitle={My title},    
    pdfauthor={Author},     
    pdfsubject={Subject},   
    pdfcreator={Creator},   
    pdfproducer={Producer}, 
    pdfkeywords={keyword1, key2, key3}, 
    pdfnewwindow=true,      
    colorlinks=true,       
    filecolor=magenta,      
    urlcolor=black           
}

\usepackage{pdfrender}

\usepackage[skins]{tcolorbox}
\newcommand{\boxSep}{0pt}
\newcommand{\boxRule}{0.5pt}
\newcommand{\boxTopLine}{2pt}
\newcommand{\boxBottomLine}{0pt}
\newcommand{\boxLeftLine}{1.2pt}
\newcommand{\boxRightLine}{1.2pt}
\newcommand{\boxArc}{2pt}

\newlength{\dpcircle}
\newlength{\rcircle}
\newlength{\dcircle}

\definecolor{label_a}{HTML}{fad4d3}
\definecolor{label_b}{HTML}{ffaaaa}
\definecolor{label_c}{HTML}{ffff00}
\definecolor{label_d}{HTML}{c1ddb7}
\definecolor{label_e}{HTML}{a7c2ab}
\definecolor{label_f}{HTML}{c0c0ff}
\definecolor{label_g}{HTML}{adedff}
\definecolor{label_h}{HTML}{e1c9e7}

\newtcbox{\rectLabelA}{enhanced,nobeforeafter,tcbox raise base,boxrule=\boxRule,top=\boxTopLine,bottom=\boxBottomLine,
  right=\boxRightLine,left=\boxLeftLine,arc=\boxArc,boxsep=\boxSep,before upper={\vphantom{dlg}},
  colframe=label_a,coltext=black,colback=label_a}
\robustify{\rectLabelA}
\newtcbox{\rectLabelB}{enhanced,nobeforeafter,tcbox raise base,boxrule=\boxRule,top=\boxTopLine,bottom=\boxBottomLine,
  right=\boxRightLine,left=\boxLeftLine,arc=\boxArc,boxsep=\boxSep,before upper={\vphantom{dlg}},
  colframe=label_b,coltext=black,colback=label_b}
\robustify{\rectLabelB}
\newtcbox{\rectLabelC}{enhanced,nobeforeafter,tcbox raise base,boxrule=\boxRule,top=\boxTopLine,bottom=\boxBottomLine,
  right=\boxRightLine,left=\boxLeftLine,arc=\boxArc,boxsep=\boxSep,before upper={\vphantom{dlg}},
  colframe=label_c,coltext=black,colback=label_c}
\robustify{\rectLabelC}
\newtcbox{\rectLabelD}{enhanced,nobeforeafter,tcbox raise base,boxrule=\boxRule,top=\boxTopLine,bottom=\boxBottomLine,
  right=\boxRightLine,left=\boxLeftLine,arc=\boxArc,boxsep=\boxSep,before upper={\vphantom{dlg}},
  colframe=label_d,coltext=black,colback=label_d}
\robustify{\rectLabelD}
\newtcbox{\rectLabelE}{enhanced,nobeforeafter,tcbox raise base,boxrule=\boxRule,top=\boxTopLine,bottom=\boxBottomLine,
  right=\boxRightLine,left=\boxLeftLine,arc=\boxArc,boxsep=\boxSep,before upper={\vphantom{dlg}},
  colframe=label_e,coltext=black,colback=label_e}
\robustify{\rectLabelE}
\newtcbox{\rectLabelF}{enhanced,nobeforeafter,tcbox raise base,boxrule=\boxRule,top=\boxTopLine,bottom=\boxBottomLine,
  right=\boxRightLine,left=\boxLeftLine,arc=\boxArc,boxsep=\boxSep,before upper={\vphantom{dlg}},
  colframe=label_f,coltext=black,colback=label_f}
\robustify{\rectLabelF}
\newtcbox{\rectLabelG}{enhanced,nobeforeafter,tcbox raise base,boxrule=\boxRule,top=\boxTopLine,bottom=\boxBottomLine,
  right=\boxRightLine,left=\boxLeftLine,arc=\boxArc,boxsep=\boxSep,before upper={\vphantom{dlg}},
  colframe=label_g,coltext=black,colback=label_g}
\robustify{\rectLabelG}
\newtcbox{\rectLabelH}{enhanced,nobeforeafter,tcbox raise base,boxrule=\boxRule,top=\boxTopLine,bottom=\boxBottomLine,
  right=\boxRightLine,left=\boxLeftLine,arc=\boxArc,boxsep=\boxSep,before upper={\vphantom{dlg}},
  colframe=label_h,coltext=black,colback=label_h}
\robustify{\rectLabelH}
\newcommand{\aLabel}{\rectLabelB{A}}
\newcommand{\bLabel}{\rectLabelD{B}}
\newcommand{\cLabel}{\rectLabelC{C}}
\newcommand{\dLabel}{\rectLabelG{D}}
\newcommand{\eLabel}{\rectLabelH{E}}

\newtcbox{\rectBlkWhtBtn}{enhanced,nobeforeafter,tcbox raise base,boxrule=\boxRule,top=\boxTopLine,bottom=\boxBottomLine,
  right=\boxRightLine,left=\boxLeftLine,arc=\boxArc,boxsep=\boxSep,before upper={\vphantom{dlg}},
  colframe=black,coltext=black,colback=white}
\robustify{\rectBlkWhtBtn}
\definecolor{stroke}{rgb}{0.2, 0.2, 0.2}

\newtcbox{\rectC}{enhanced,nobeforeafter,tcbox raise base,boxrule=\boxRule,top=\boxTopLine,bottom=\boxBottomLine,
  right=\boxRightLine,left=\boxLeftLine,arc=\boxArc,boxsep=\boxSep,before upper={\vphantom{dlg}},
  colframe=pitch_c,coltext=black,colback=white}
\newtcbox{\rectD}{enhanced,nobeforeafter,tcbox raise base,boxrule=\boxRule,top=\boxTopLine,bottom=\boxBottomLine,
right=\boxRightLine,left=\boxLeftLine,arc=\boxArc,boxsep=\boxSep,before upper={\vphantom{dlg}},
colframe=pitch_d,coltext=black,colback=white}
\newtcbox{\rectE}{enhanced,nobeforeafter,tcbox raise base,boxrule=\boxRule,top=\boxTopLine,bottom=\boxBottomLine,
right=\boxRightLine,left=\boxLeftLine,arc=\boxArc,boxsep=\boxSep,before upper={\vphantom{dlg}},
colframe=pitch_e,coltext=black,colback=white}
\newtcbox{\rectF}{enhanced,nobeforeafter,tcbox raise base,boxrule=\boxRule,top=\boxTopLine,bottom=\boxBottomLine,
right=\boxRightLine,left=\boxLeftLine,arc=\boxArc,boxsep=\boxSep,before upper={\vphantom{dlg}},
colframe=pitch_f,coltext=black,colback=white}
\newtcbox{\rectG}{enhanced,nobeforeafter,tcbox raise base,boxrule=\boxRule,top=\boxTopLine,bottom=\boxBottomLine,
right=\boxRightLine,left=\boxLeftLine,arc=\boxArc,boxsep=\boxSep,before upper={\vphantom{dlg}},
colframe=pitch_g,coltext=black,colback=white}
\newtcbox{\rectA}{enhanced,nobeforeafter,tcbox raise base,boxrule=\boxRule,top=\boxTopLine,bottom=\boxBottomLine,
right=\boxRightLine,left=\boxLeftLine,arc=\boxArc,boxsep=\boxSep,before upper={\vphantom{dlg}},
colframe=pitch_a,coltext=black,colback=white}
\newtcbox{\rectB}{enhanced,nobeforeafter,tcbox raise base,boxrule=\boxRule,top=\boxTopLine,bottom=\boxBottomLine,
right=\boxRightLine,left=\boxLeftLine,arc=\boxArc,boxsep=\boxSep,before upper={\vphantom{dlg}},
colframe=pitch_b,coltext=black,colback=white}
\newtcbox{\rectFis}{enhanced,nobeforeafter,tcbox raise base,boxrule=\boxRule,top=\boxTopLine,bottom=\boxBottomLine,
right=\boxRightLine,left=\boxLeftLine,arc=\boxArc,boxsep=\boxSep,before upper={\vphantom{dlg}},
colframe=pitch_fis,coltext=black,colback=white}
\newtcbox{\rectCis}{enhanced,nobeforeafter,tcbox raise base,boxrule=\boxRule,top=\boxTopLine,bottom=\boxBottomLine,
right=\boxRightLine,left=\boxLeftLine,arc=\boxArc,boxsep=\boxSep,before upper={\vphantom{dlg}},
colframe=pitch_cis,coltext=black,colback=white}
\newtcbox{\rectGis}{enhanced,nobeforeafter,tcbox raise base,boxrule=\boxRule,top=\boxTopLine,bottom=\boxBottomLine,
right=\boxRightLine,left=\boxLeftLine,arc=\boxArc,boxsep=\boxSep,before upper={\vphantom{dlg}},
colframe=pitch_gis,coltext=black,colback=white}

\robustify{\rectC}
\robustify{\rectD}
\robustify{\rectE}
\robustify{\rectF}
\robustify{\rectG}
\robustify{\rectA}
\robustify{\rectB}
\robustify{\rectFis}
\robustify{\rectCis}
\robustify{\rectGis}

\usepackage{microtype}
\usepackage{calc}
\newlength\myheight
\newlength\mydepth
\settototalheight\myheight{Xygp}
\PassOptionsToPackage{svgnames, x11names, dvipsnames}{xcolor}

\newtcbox{\inlinebox}[1][]{box align=bottom,
 nobeforeafter,
 colback=#1,
 size=small,
 grow to left by=-1pt,
 grow to right by=-1pt,
 boxrule=0pt,
 sharp corners}
 \newtcbox{\inlineboxwithtext}[1][]{tcbox raise=-5pt,
 nobeforeafter,
 colback=#1!20,
 colframe=#1!85!black,
 size=fbox,
 boxrule=1pt,
 enlarge by=1pt,
 enlarge top by=0pt,
 sharp corners}

\newcommand{\subhead}[1]{\vspace{2pt} \noindent \textbf{#1}}
\newcommand{\subheadnvs}[1]{\noindent \textbf{#1}}

\definecolor{label_a}{HTML}{fad4d3}
\definecolor{label_f}{HTML}{c0c0ff}
\definecolor{label_g}{HTML}{adedff}
\definecolor{label_h}{HTML}{e1c9e7}

\newcommand{\xtramargin}[1]{#1}
\newcommand{\appurl}{https://visual-musicology.com/corpus}
\newcommand{\corpusvis}{\href{\appurl}{CorpusVis}}

\title[CorpusVis: Visual Analysis of Digital Sheet Music Collections]{CorpusVis: Visual Analysis of Digital Sheet Music Collections}

\author[
M. Miller, J. Rauscher, D. Keim \& M. El-Assady
]
{
    \parbox{\textwidth}{\centering
    \vspace{-20pt} 
    Matthias Miller$^{1}$\orcid{0000-0002-6281-2173},
    Julius Rauscher$^{1}$,\orcid{0000-0003-1318-9642},
    Daniel A. Keim$^{1}$\orcid{0000-0001-7966-9740}, 
    and Mennatallah El-Assady$^{2}$\orcid{0000-0001-8526-2613}
    }
    \\
    {\parbox{\textwidth}{\centering 
        \vspace*{-5pt} $^1$University of Konstanz, Germany \hspace{1cm}
        $^2$ETH AI Center, Zürich, Switzerland
    }}
}

\begin{document}

\teaser{
  \vspace*{-2em}
  \centering
  \href{\appurl?arxiv=true}{\includegraphics[width=1\textwidth]{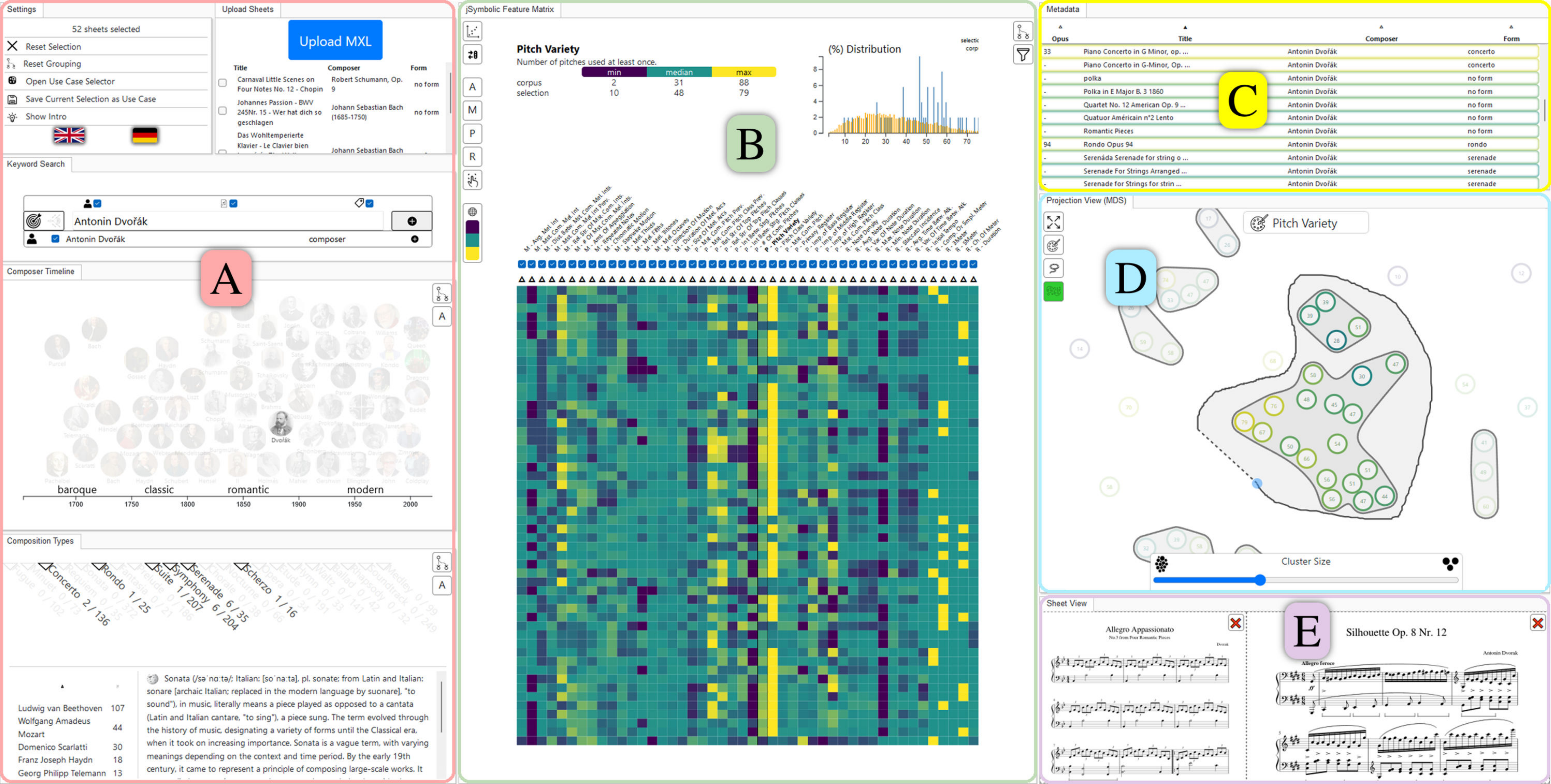}}
  \caption{
  Our visual interactive workspace supports the analysis of sheet music collections through visualizations that are connected through linking and brushing. 
  After filtering a sheet music corpus~\aLabel, music analysts can perform exploration, comparison, and detection tasks. 
  The feature matrix~\bLabel~provides detailed information about low-level characteristics for musical compositions. 
  The metadata table~\cLabel~enables browsing through the titles, composers, and composition forms.
  An MDS projection view~\dLabel~helps to identify similar compositions or even clusters based on selected features.
  The sheet view~\eLabel~allows analysts to view compositions using the familiar notation.
  Domain experts can use \corpusvis~to confirm and generate hypotheses and detect interesting patterns between composers and composition types. 
  }
  \label{fig:teaser}
}

\maketitle

\begin{abstract}
    Manually investigating sheet music collections is challenging for music analysts due to the magnitude and complexity of underlying features, structures, and contextual information. 
    However, applying sophisticated algorithmic methods would require advanced technical expertise that analysts do not necessarily have. 
    Bridging this gap, we contribute \corpusvis, an interactive visual workspace, enabling scalable and multi-faceted analysis. 
    Our proposed visual analytics dashboard provides access to computational methods, generating varying perspectives on the same data.
    The proposed application uses metadata including composers, type, epoch, and low-level features, such as pitch, melody, and rhythm.
    To evaluate our approach, we conducted a pair-analytics study with nine participants. 
    The qualitative results show that CorpusVis supports users in performing exploratory and confirmatory analysis, leading them to new insights and findings. 
    In addition, based on three exemplary workflows, we demonstrate how to apply our approach to different tasks, such as exploring musical features or comparing composers.
\end{abstract}
\section{Introduction}

Investigating music collections are relevant for both domain experts such as musicologists and music theorists and usual music consumers~\cite{weihs2016music}. Predominantly, the audio format has been the starting point for many applications including music classification~\cite{staraudiobasedmusicclassification2011fu}, personalized music recommendation~\cite{semanticaudiomusicrecomendation2013bogdanov}, structure analysis~\cite{star_audiobasedstructureanalysis2010paulus}, and the generation of music playlists~\cite{playlistgeneration2021sakurai}. In contrast, we are convinced that the investigation of sheet music collections based on digital symbolic formats such as MusicXML~\cite{good2001musicxml_virtualscore}, MIDI, Humdrum~\cite{sapp2015graphic}, and Lilypond~\cite{nienhuys2003lilypond} have received less attention compared to audio.

While services such as sheet music recommendation offered by, e.g., OKTAV~\cite{oktav2021sheetrecommendation} leverage musical features and individual preferences to provide sheet music suggestions for piano players, it remains unclear how such recommendations are computed. To understand which musical features (e.g., structure or melody) reveal differences and similarities, computational approaches for sheet music analysis have been proposed in the community, supporting various tasks such as genre and composer classification~\cite{genreclassificationsymbolic2017armentano}. Besides close reading of sheet music as a typical task in musicology, distant reading is also an essential task in the digital humanities~\cite{CloseDistant2015Jaenicke}. Close reading supports musicologists to investigate musical compositions on a detailed level, abstract visualization contain the potential to provide an overview over a large set of musical compositions within their context, thus enabling distant reading. While approaches that solely rely on algorithmic or computational approaches mitigate the tedium of manual analysis of a dataset done by human analysts, they often are not accessible to a broader user group. Still, they are limited to music analysts that have programming skills. To address this gap, tailored visualizations combined with user interaction can be employed to increase access to music analysis methods for a broader user group while taking the human analyst in the analysis loop, which is a fundamental aspect for knowledge generation in visual analytics~\cite{sacha2014knowledgegeneration}. An essential aspect of the understanding process is critical thinking by the analyst about the subject, which was introduced by Bradley et al. as \emph{slow analytics}~\cite{bradley2018visualization}.

The availability of sheet music datasets such as KernScores~\cite{DBLP:conf/ismir/Sapp05} or MuseScore~\cite{musescorecolunteered2016} have the potential for analysis at different scales, benefiting from using projection techniques based on underlying features~\cite{jsymbolic22mckay2018}. While MuseScore contains more than a million compositions, the quality of the uploaded material varies from content that is provided by users who ensure that provided content is faithful to the original compositions to pieces that suffer from inaccuracy or incompleteness. Therefore, it is challenging to maintain a high quality of the underlying data for the analysis when considering the full dataset, requiring expensive data cleaning steps before performing analysis tasks on it. Manual curation or selection processes are required to set the focus on a particular subset that is either provided by users, that only provides high-quality content. Alternatively, analysts can manually view each composition that shall be part of a curated dataset, which is, of course, quite a tedious process. The research field of optical music recognition provide approaches to convert printed sheet music into symbolic formats such as MusicXML. Yet, it needs further improvements until musicologists trust these automatic results~\cite{omr_star2020shatri}. Enabling analysts to influence the data foundation is crucial for the analysis.

Besides data quantity, increasing the quality of the data is crucial for effective analysis. This includes a targeted selection of representative samples. For example, a basic issue in humanities research is striving to  avoid the exclusion of marginalized positions, a typical risk of data colonialism. For instance, famous classical composers such as Johannes Sebastian Bach, Wolfgang Amadeus Mozart, Joseph Haydn, or Ludwig van Beethoven are often a more prominent analysis subject in musicology  than less known composers such as Muzio Clementi, Domenico Scarlatti, or Francois-Joseph Gossec. This marginalization of composers is reaffirmed by the vast amount and duplicates of famous compositions in datasets such as MuseScore. To enable a large-scale analysis of sheet music collections, a corpus-level overview is essential. This has the power to enable analysts to audit and refine corpora, as well as compare music sheets on numerous features. Applying abstract visualization techniques to sheet music corpora at a larger scale has the potential to provide insights about the work of composers or typical differences between compositions types without the need to manually analyze every detail of single data items.

Driven by the need for a corpus-level music analysis technique, in this work, we address the research question: 
\textit{How to support music analysts to explore, investigate, and compare sheet music collections based on metadata and low-level features using interactive visualization?} Our aim is to provide them with a multiscale and multi-perspective bird's-eye view on sheet music corpora. We thus designed an interactive visual workspace that provides multiple tailored analysis components in an inter-linked dashboard.  The visual analysis is supported by computational methods, such as clustering, to aid in pattern finding. Analysts can openly explore the underlying corpus, investigate a set of pre-configured use cases,  or verify hypotheses through crafting their own analysis workflow.

\subhead{Contributions -- }
This work contributes a problem characterization addressing visualization requirements with regard to the analysis of sheet music corpora. We provide a list of relevant data and task characteristics, as well as a description of target audiences. 
A major contribution is the Visual Analysis Workspace for sheet music collections through a combination of multiple components that are seamlessly connected through linking and brushing. We conducted a qualitative evaluation to assess the applicability of our approach and provide details about its benefits and drawbacks. Finally, we discuss open research opportunities to inspire interdisciplinary collaboration at the interface of visualization and musicology.

\vspace*{-10pt}
\section{Related Work} \label{sec:related-work}
The field of musicology is a wide research area that covers heterogeneous research questions and challenges. The subfield \emph{Visual Musicology} at the interface of musicology and visualization research as introduced by Miller et al. illustrates the vast opportunities for which visualization could be applied to support domain-related, scientific issues~\cite{framingvisualmusicology2019}. Their framework highlights the potential of interactive visualization to perform analysis tasks, including information retrieval, exploration, and comparison. We use the \href{https://visual-musicology.com/graph/}{\emph{visual musicology graph}} to classify the work presented in this paper accordingly. Specifically, we focus on \emph{Structural Features} and \emph{Meta-Information} of sheet music and the visualization tasks \emph{Overview/Summarization}, \emph{Navigation/Exploration}, \emph{Clustering}, \emph{Comparison}, and \emph{Details on Demand} within the domains \emph{Theory \& Analysis} and \emph{History}. 

\subhead{Visual Analysis of Music Collections} -- 
The visual investigation of musical data collections has already received attention by visualization researchers. Khulusi et al. created a web-based interface called ``musiXplora'' as part of a digitization project that enables humanities researchers to investigate musicological data such as meta-information (e.g., gender, religious denomination, profession, institutions) of musicians and instrument makers~\cite{musixplora2020khulusi}. For instance, this interactive prototype facilitates information retrieval and browsing about the life and work of different composers. Similarly, Jänicke et al. designed an interactive visualization tool for the interactive profiling of musicians and their relationships based on their meta-information~\cite{interactiveprofiling2016jaenicke}.  They did not consider specific features about the compositional work but rather focus on the contextual information about the life, profession, and instruments they mastered. Pampalk used SOMs to visualize and classify the genre of music data as ``Islands of Music'' based on audio-extracted features such as the loudness information from audio files~\cite{islandsofmusic2001pampalk}. Chen and Putz designed an interactive UI for browsing and organizing music collection allowing to listen and visually explore the similarity of musical pieces based on low-level features that reflect the genre and style of pieces from different composers~\cite{MusicSimChen09}. Weiß et al. show how exact tonal features (pitch information) and metadata from audio data help to detect style changes over the different epochs from baroque to modern music~\cite{weiss2018}. Georges and Nguyen use a dataset of 500 classical composers to visually analyze different epochs using dendrograms and MDS projection techniques~\cite{georges2019mds_musicsimilarity}. By that, they demonstrate how abstract visualization methods support the comparison of composers~\cite{slowanalytics2016bradley}. 

\subhead{Feature Extraction from Music Data --} 
Music is available and can be stored based on different formats such as symbolic (e.g., MusicXML~\cite{good2001musicxml_virtualscore}) or audio data (e.g., mp3). Depending on this format, different features can be extracted that can afterwards be used for different tasks such as music recommendation~\cite{schedl2019deep}. Similar to extracting features from audio signal data (e.g., spectral information) for further processing~\cite{surfboardfeatures2020lenain}, it is possible to extract low-level features from symbolic music~\cite{music21jsymbolicfeatures}. For example, \href{https://web.mit.edu/music21/}{music21}~\cite{music21jsymbolicfeatures} is a Python library that enables programmers to extract jSymbolic features from different symbolic sheet music formats including MusicXML. McKay and Fujinaga discuss how such features can be used for MIR research tasks~\cite{jsymbolic22mckay2018} such as composer classification as done by Verma and Thickstun leveraging convolutional networks~\cite{convo_composer_class2019verma}. McKay leverages audio and symbolic music data sources to perform music classification and the creation of music information retrieval tools using his own jMIR software suite that can even consider contextual/cultural metadata~\cite{DBLP:conf/icmc/McKayF09}. Corr{\^{e}}a and Rodrigues published a survey about music genre classification that reveals that such tasks primarily use automatic algorithms without using advanced visualization techniques~\cite{DBLP:journals/eswa/CorreaR16}. Merely computational approaches do not enable music analysts to explore low-level features, making it impossible to step into the analysis process to get a better overview over the inner workings of the applied algorithms.

\subhead{Visualization of Collections in the Digital Humanities --}
Besides musicology, there exist other scientific issues within digital humanities that have already been addressed by information visualization researchers such as the visual analysis of poems~\cite{poemage16mccurdyLCM}. Jänicke et al. discuss many digital humanities projects that require text processing methods~\cite{JanickeFCS17visualtextanalysis} while Kirschenbaum also confirms the essentiality of text for humanities research ranging from close via not-reading to distant reading~\cite{kirschenbaum2007remaking}. For example, Bludeau et al. implemented a web-based prototype to enable the visual investigation of literature and handwritten notes from \emph{Fontane's Handbibliothek}~\cite{readingtraces2020bludeau}. For instance, the field of text analysis deals with comparison tasks for plagiarism detection by identifying typical features for specific authors~\cite{GippMBPN14visualplagiarismdetection}. Keim and Oelke use textual features to create a literature fingerprinting visualization that facilitates the analysis of whole books~\cite{keimO07literaturefingerprinting}. We argue that existing visualization methods that have been successfully applied to similar endeavors in related research areas enclose the potential for methodology transfer~\cite{framingvisualmusicology2019}. Consequently, we consider it to be useful to get inspired by existing visual techniques to address unsolved challenges such as the visual analysis of sheet music collections.

\subhead{Research Gap --} 
While there are several projects within this research field, they are often limited to audio data only~\cite{MusicSimChen09,weiss2018}. Thus, we argue that the analysis of music based on symbolic formats has received less attention. We argue this shift towards audio data could be due to larger user groups of music applications such as (e.g., Spotify) which benefit from effective music recommendation methods. We see the reason for this unbalance in the smaller group size of sheet music consumers, which is much smaller. Liem et al. discuss the issue of focusing on audio signal only and argue for multimodal and user-centered analysis strategies to improve the accessibility of digital music data~\cite{liem_need_2011,multimodalproc2011mueller}. Nevertheless, the increased use of digital symbolic music formats led to services such as OKTAV~\cite{oktav2021sheetrecommendation} which provides sheet music recommendations based on symbolic features. While automatic solutions have their application areas, employing visual methods allows analysts to step into the analysis process, helping them to gain new insights or generate and confirm hypotheses, which is also known as human-in-the-loop approaches~\cite{sacha2014knowledgegeneration}. To our knowledge, there is no previous work that readily supports the analysis of sheet music corpora, facilitating the visual comparison of metadata and low-level features.
\newcommand{\visA}{\textbf{\texttt{[D1]}}}
\newcommand{\visB}{\textbf{\texttt{[D2]}}}
\newcommand{\visC}{\textbf{\texttt{[D3]}}}

\newcommand{\taskA}{\textbf{\texttt{[T1]}}}
\newcommand{\taskB}{\textbf{\texttt{[T2]}}}
\newcommand{\taskC}{\textbf{\texttt{[T3]}}}
\newcommand{\taskD}{\textbf{\texttt{[T4]}}}
\newcommand{\taskE}{\textbf{\texttt{[T5]}}}
\newcommand{\taskF}{\textbf{\texttt{[T6]}}}
\newcommand{\taskG}{\textbf{\texttt{[T7]}}}

\section{Problem Description}
\label{sec:problemdescription}
Our primary objective is to provide an \emph{interactive visual workspace} to support the investigation and exploration of sheet music collections. The target audience primarily comprises musicologists and music theorists. We assume that music librarians and Music Information Retrieval (MIR) researchers could benefit from our work as well. For this reason, we gathered information about the data and task requirements by conducting initial expert interviews and consulted existing literature. 

\subhead{Feature Characteristics --} 
When analyzing musical compositions, it is essential to musicologists to know the relationship to the \emph{composers} which allows to include domain knowledge into the analysis process. Usually, one can assign musical pieces to a certain \emph{type or style} which often correlates to a specific genre. Considering this contextual information can help detect and compare entities from a musical score collection. As discussed in~\autoref{sec:related-work}, there exist various frameworks that allow for extracting \emph{statistical features} from symbolic music. We use music21 to extract jSymbolic features from digital sheet music~\cite{Cuthbert2010} that provides the foundation for the visual analysis workspace that we introduce in the following sections. Specifically, these low-level features contain rhythmic, melodic, and pitch information that provide informative attributes about single compositions or numerical distributions when investigating a score collection. In addition, we consider available contextual metadata including composer, composition type, and temporal characteristics.

\subhead{Task Characteristics --}
Content-based MIR aims at uncovering music data relevant characteristics that can be employed for similarity computations and retrieval tasks~\cite{georges2019mds_musicsimilarity}. For instance, a specific retrieval task \taskA would be the identification of all compositions from \textit{Mozart} that have a time signature of a \textit{triple meter}. Besides performing inquiring based on exact information, music analysts could also be interested in exploring similarities of pieces within a dataset \taskB with the objective to create new hypotheses or gain new insights about the underlying information~\cite{Khulusi2020}. Additionally, the comparison of two or even a set of compositions \taskC can be relevant during music analysis to identify commonalities of the complete work of composers or different versions of the same composition~\cite{DBLP:conf/cmmr/UrbanoLMS10}. Not only comparison tasks can benefit from grouping and clustering based on meta information \taskD, but also the investigation of single composers based on statistical values, they provide a quick overview without the need to view all items separately. Depending on the quality of the dataset, it can be crucial for analysts to identify duplicates within a dataset \taskE to omit works from further analysis or other issues such as detecting musical plagiarism~\cite{Cameron2020Plagiarism}. Based on our cooperation with musicologists we focus on the following tasks that address typical analysis challenges:\\[2pt]
\subheadnvs{\taskA}~Retrieve/Filter compositions based on title, composer, or type \\
\subheadnvs{\taskB}~Exploratory analysis of a musical score collection \\
\subheadnvs{\taskC}~Comparison of scores, composers, types, and features \\
\subheadnvs{\taskD}~Clustering/grouping of pieces, composers, and types \\
\subheadnvs{\taskE}~Detection of similar entities in a given corpus

\subhead{Visualization Requirements --} 
Based on the described data and task characteristics, we can identify essential requirements for the visualizations that are necessary to support the analysis process. For the information retrieval task \taskA, we need a separate filter and sort functionality for each meta feature and the low-level characteristics. Suppose a user is looking for a specific composer or composition. In that case, we need an interactive search bar that allows for direct querying of the underlying dataset to avoid the tedious search for data items within the provided visualizations. To associate the temporal aspects of composers, we need a timeline visualization that can also be directly used to filter composers to be analyzed by the user. The type of composition is another fundamental characteristic for music analysts that can be used for exploration~\taskB, comparison~\taskC, or identification of interesting pieces. To extract similar compositions within the dataset, we can employ a dimension reduction technique such as MDS~\cite{cox2008multidimensional} to project dozens of low-level features into a two-dimensional space~\taskE. Then, identifying groups based on a clustering method is required to detect multiple pieces that have related features~\taskD.

\subhead{Used Dataset --}
MuseScore comprises a heterogeneous dataset with over a million compositions. Instead of using all these files, we manually selected a subset from certain users who upload curated files, e.g., \href{https://musescore.com/user/19710/sheetmusic}{ClassicMan}~\cite{ClassicMan2021MuseScore}. Typically, these compositions have high community ratings. Because of this preselection process, we ensure that the compositions in the corpus are of better quality.

\newcommand{\dd}{${D}$}
\newcommand{\mC}{${C}$}
\newcommand{\mF}{${F}$}
\newcommand{\mP}{${P}$}
\newcommand{\mT}{${T}$}

\begin{figure}[b]
    \centering
    \xtramargin{\vspace*{-5pt}}
    \includegraphics[width=\linewidth]{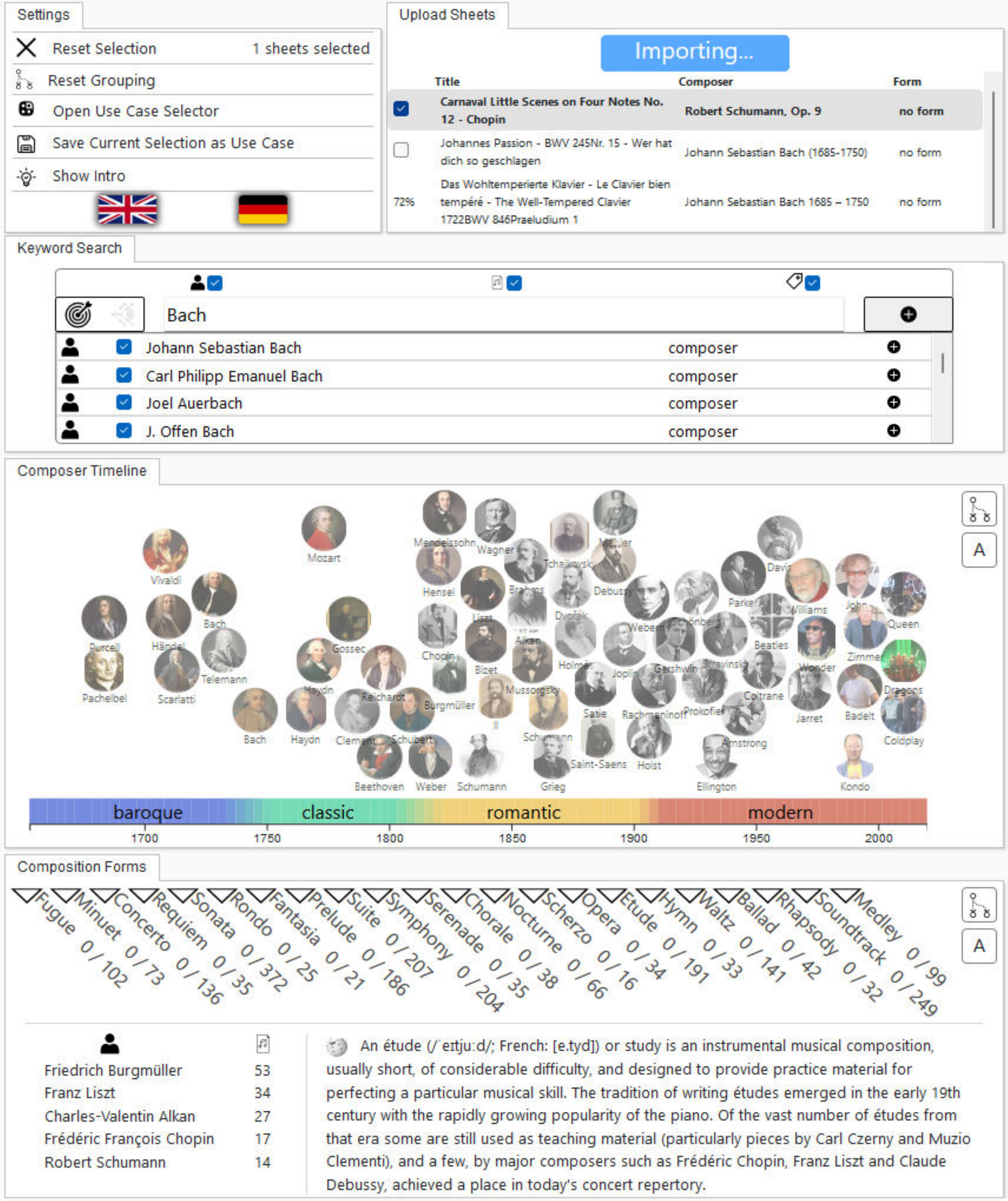}
    \xtramargin{\vspace*{-15pt}}
    \caption{
    The keyword search allows to filter composers and composition types.
    Users can extend the dataset by uploading custom data. 
    Further settings: reset the system state, save use cases, or open an interactive introductory guide through through~\corpusvis.}
    \label{fig:ws_dataSelection}
    \xtramargin{\vspace*{-5pt}}
\end{figure}

\section{Visual Interactive Analysis Workspace}
\label{sec:workspace}
This section introduces the interactive workspace and describes three exemplary analysis workflows to illustrate how to use it for analyzing sheet music corpora.
\corpusvis~is available as a web-based application via~ \href{\appurl?system=true}{\textbf{\texttt{\textcolor{blue}{\appurl}}}}.

\subsection{CorpusVis -- User Interface}
The analysis workspace consists of five components (see~\autoref{fig:teaser}) supporting data filtering and visual analysis following Shneiderman's Visual Information Seeking Mantra~\cite{shneiderman96mantra}. The components are connected through animated linking and brushing, showing relevant details during the analysis process to the user. For instance, if the user hovers a circle (represents a composition or aggregations) in the projection view~\dLabel, then the corresponding composer and forms~\aLabel~and matrix row(s) matrix~\bLabel~are highlighted. The settings widget (\autoref{fig:ws_dataSelection}, top left) enables users to switch between the German and English and provides a guided tour for all components.

\subhead{Data Import and Selection --} 
This component enables uploading custom sheet music datasets (\emph{.mxl} format) or filtering existing datasets by keywords (composer names, title, etc.), composition types, and via composer selection from the timeline~\taskA. The user can save an interesting data selection as a use case, or simply reset the analysis through further options provided in the settings widget.
The composer timeline shows 62 composers for which we could directly extract their life data from the files. Similar to Weiß et al.~\cite{weiss2018}, we use four historical periods (baroque, classic, romantic, modern) to position the composers on the timeline. We apply an animated layout that uses the vertical axis to avoid overlapping.

\begin{figure}[t]
    \centering
    \includegraphics[width=\linewidth]{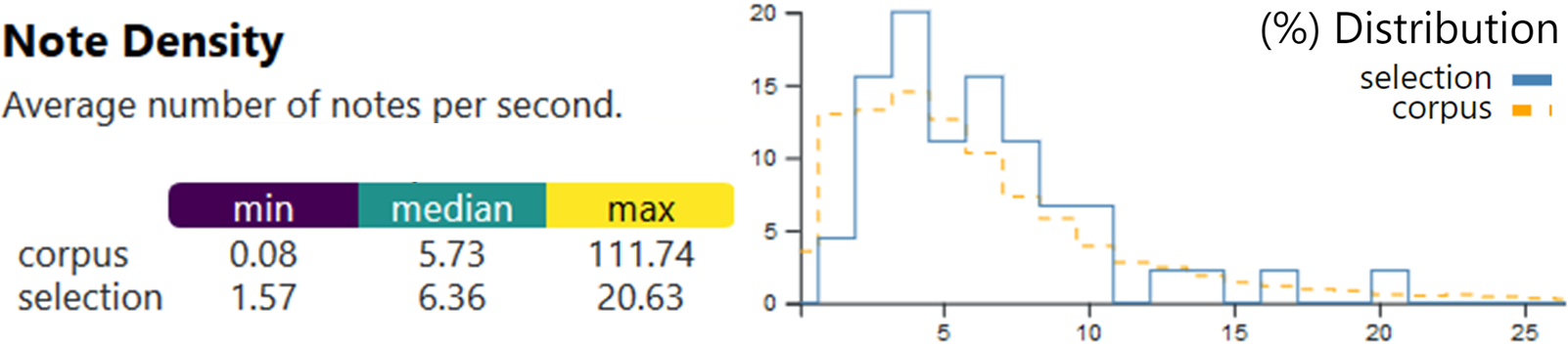}
    \xtramargin{\vspace*{-15pt}}
    \caption{ The feature distribution chart provides statistical characteristics (min., median, max.) of a subset and enables comparing the subset to the rest of the corpus regarding a selected feature. }
    \label{fig:featuredistributionchart}
    \xtramargin{\vspace*{-10pt}}
\end{figure}

\begin{figure}[b]
    \centering
    \xtramargin{\vspace*{-8pt}}
    \includegraphics[width=\linewidth]{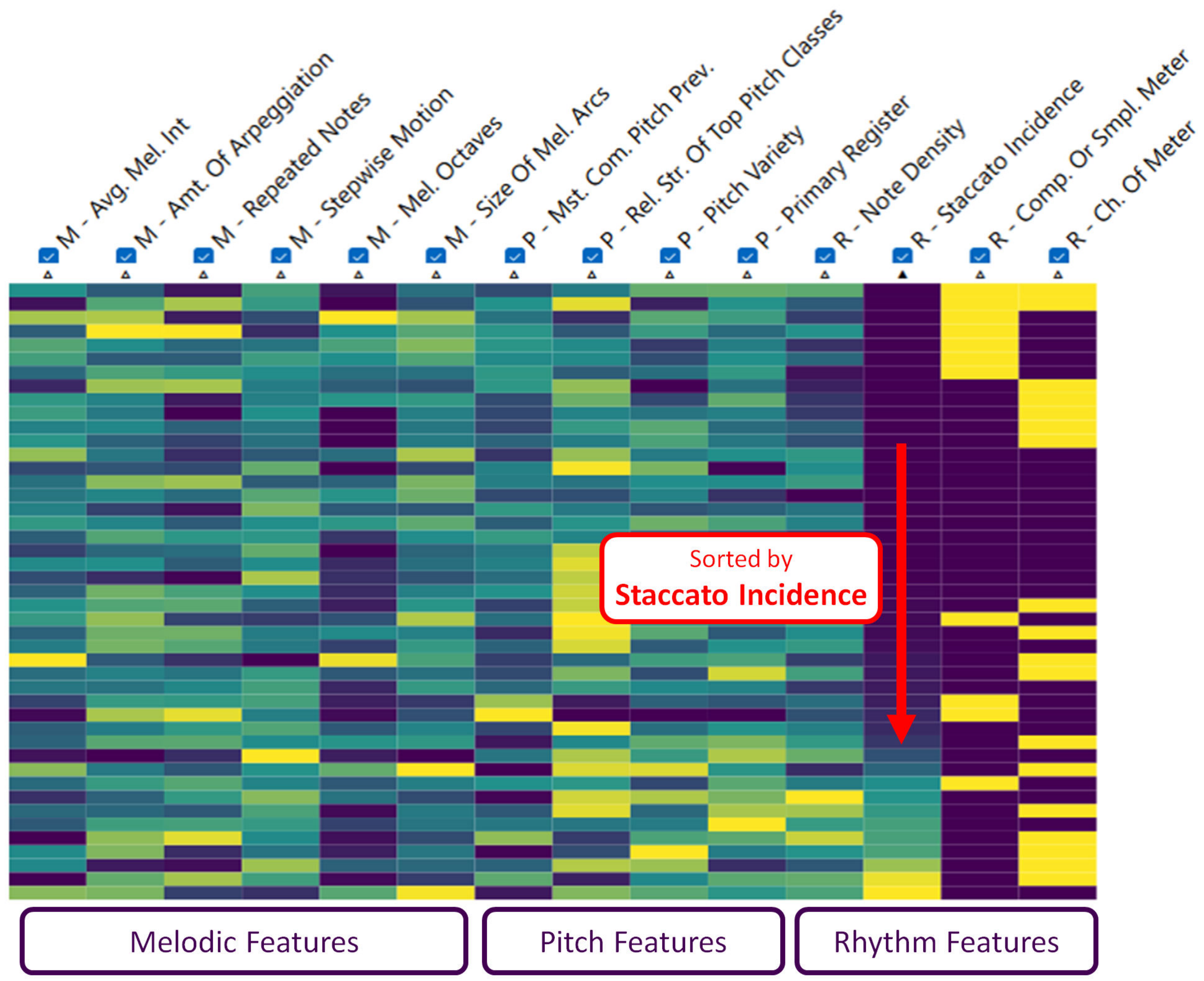}
    \xtramargin{\vspace*{-15pt}}
    \caption{This exemplary matrix shows 14 features from the two composers \textit{Clara Schumann} and \textit{Modest Petrovich Mussorgsky}. Users can visually detect low (purple) and high (yellow) values. }
    \label{fig:jsymbolicFeatureMatrix}
    \xtramargin{\vspace*{-1em}}
\end{figure}

\subhead{jSymbolic Feature Matrix --}
Depending on the grouping, the feature matrix encodes statistical information extracted from single compositions or clusters at the workspace' center. User interaction allows sorting and filtering all features to focus only on subsets if needed~\taskC.
Analysts can also inform themselves by getting detailed information for each available feature. The features support the identification of salient patterns, and the matrix can be used for comparison tasks~\taskB. Overall, the feature matrix can display up to 46 different jSymbolic features extracted with music21~\cite{music21jsymbolicfeatures}. The feature distribution chart~(see~\autoref{fig:featuredistributionchart}) helps analysts compare the statistics of a feature to the whole corpus, including the minimum, median, and maximum value of the feature for each group~\taskC. This example shows the statistical information of the feature \emph{Note Density,} which corresponds to the average number of notes that are played within a second. The values show that the corpus contains pieces with a note density above 100, while the selection is situated has higher values at the left side of the distribution compared to the overall corpus. The distribution chart illustrates that the feature regarding the selection is similar to the rest of the corpus (left-skewed). \autoref{fig:jsymbolicFeatureMatrix}~displays 14 features horizontally sorted by category (M, P, R) with the rhythmic feature \emph{Staccato Incidence} emphasizing a few pieces with a high amount of staccato usage.

\begin{figure}[t]
    \centering
    \href{https://visual-musicology.com/corpus?usecase=tonality-vs-atonality}{\includegraphics[width=\linewidth]{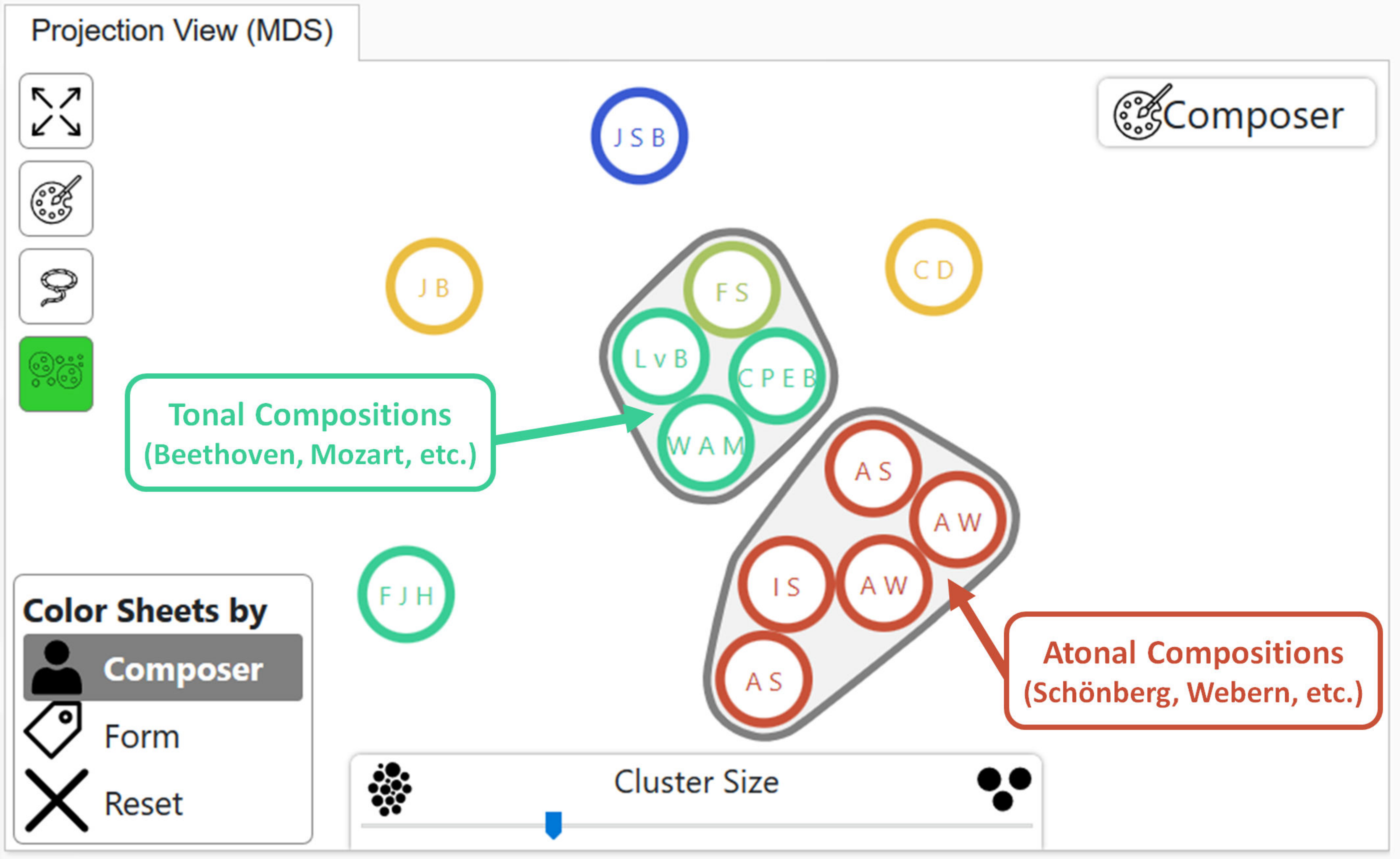}
    }
    \xtramargin{\vspace*{-15pt}}
    \caption{ MDS Projection View: If no grouping is applied, each circle in the projection view represents a single composition. The color provides the information to which epoch a piece is assigned. }
    \label{fig:projection_view}
    \xtramargin{\vspace*{-15pt}}
\end{figure}

\subhead{MDS Projection View --} We compute a multidimensional scaling projection based on the selection of low-level features to show the similarity between single compositions or clusters in a separate component~\taskE. 
The circle labels displayed in \autoref{fig:projection_view} are the initials of the composer names. In this case, the color of the circles provides epoch information, and the respective color scale is shown in the composer timeline. Analogously, a separate color scheme is used for the composition types. On-demand, the user can execute a DBSCAN clustering~\cite{HinneburgK98dbscan} and manipulate its \textit{EPS} parameter via the cluster slide at the bottom to identify groups of similar entities~\taskD. DBSCAN has the advantages that it does not require a parameter for the number of clusters and it can identify outliers. We set $\mathrm{MinPTS} = 2$ to enable the identification of duplicates. The visual grouping of the clusters uses concaveman~\cite{concaveman2017javascript}, a JavaScript library that builds on an algorithm based on work by Park and Oh~\cite{DBLP:journals/jise/ParkO12} which allows to create concave hulls around given clusters.

\begin{figure}[ht]  
    \xtramargin{\vspace*{-5pt}}
    \centering
    \includegraphics[width=\linewidth]{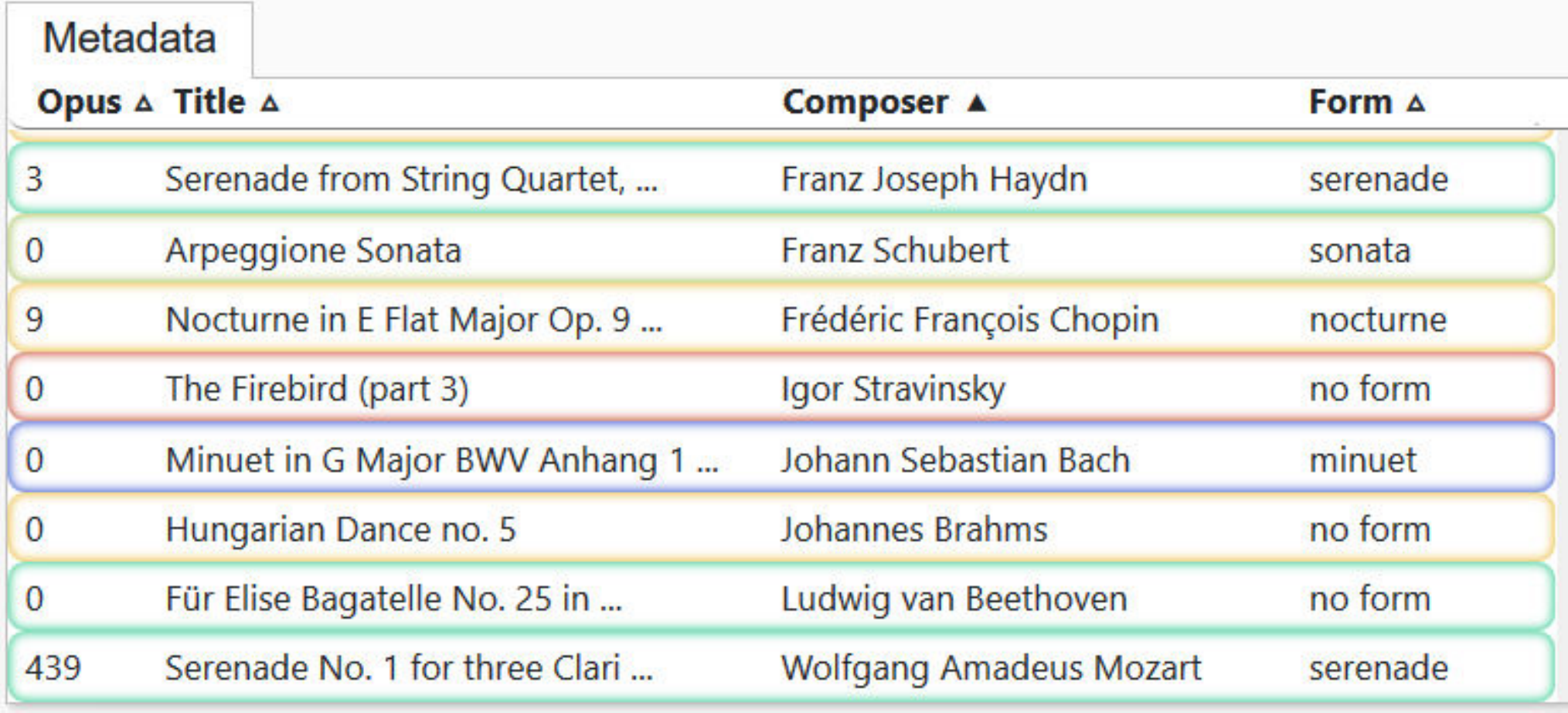}
    \xtramargin{\vspace*{-15pt}}
    \caption{Analysts can use the metadata table to sort selected compositions and simply lookup interesting pieces from the list.}
    \label{fig:metadata_table}
    \xtramargin{\vspace*{-4pt}}
\end{figure}

\subhead{Metadata Table --} As a quick reference to the feature matrix, the workspace contains a meta information table of the current selection as displayed in~\autoref{fig:metadata_table}. Users can alphabetically sort a current selection based on composers, types, titles, and opus numbers. This rather simple data representation facilitates the connection to the rows of the more abstract feature matrix (see~\autoref{fig:jsymbolicFeatureMatrix}) and projection view (see~\autoref{fig:projection_view}).

\begin{figure}[t]
    \centering
    \includegraphics[width=\linewidth]{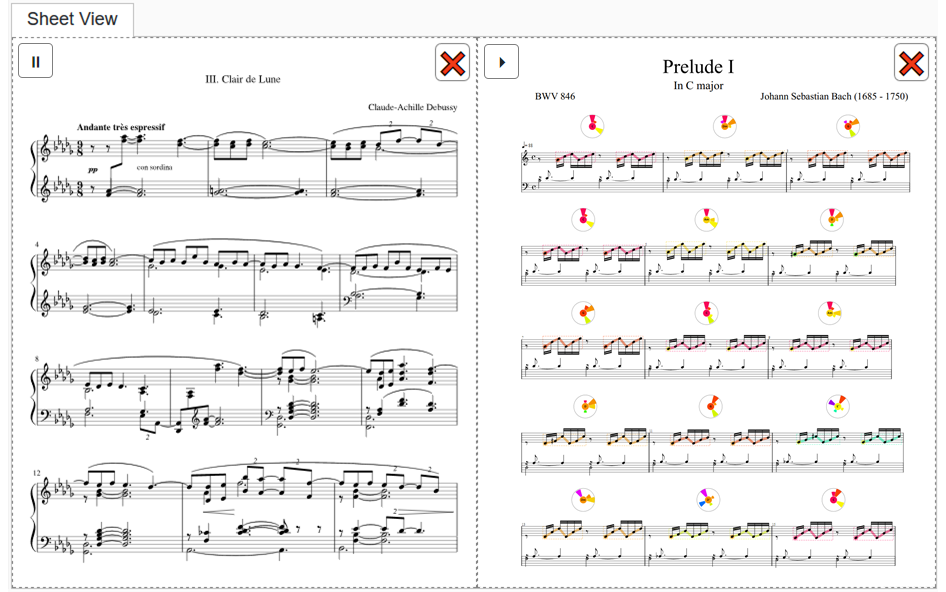}
    \xtramargin{\vspace*{-15pt}}
    \caption{
    The sheet view allows viewing the first page of compositions. 
    At the sheet level, analysts can perform harmonic and rhythmic analysis tasks using \href{https://visual-musicology.com/sheetmusicvis/?music-sheet=8}{\textbf{MusicVis}}~\cite{MiFuHa2021Augmenting}.}
    \label{fig:sheetview}
    \xtramargin{\vspace*{-15pt}}
\end{figure}

\subhead{Sheet View --} Viewing sheet music using standard visual encoding is crucial for music analysts. Thus, the last component is the \emph{sheet view,} which music analysts can use to estimate the quality of the underlying content (see \autoref{fig:sheetview}). Users can open the first page of compositions and relate it to the features via the context menu. Hence, the analyst can better understand abstract features by seeing the direct relation through the familiar notation. Via the playback icons at the top left corner of each item in the sheet view, analysts can also listen to the compositions on demand. We use the command-line version of MuseScore~\cite{musescorecolunteered2016} to convert the underlying MusicXML files into their .mp3 version. These files are also cached to reduce loading times if multiple users want to play back the same compositions. 

\subsection{Analysis Use Cases}
\label{sec:workflows}
\newcommand{\epochcomparison}{https://visual-musicology.com/corpus?usecase=epoch-comparison}
\newcommand{\composercomparison}{https://visual-musicology.com/corpus?usecase=composer-comparison}
\newcommand{\featureinvestigation}{https://visual-musicology.com/corpus?usecase=feature-investigation}
\newcommand{\atonality}{https://visual-musicology.com/corpus?usecase=tonality-vs-atonality}
\newcommand{\composeroverview}{https://visual-musicology.com/corpus?usecase=composer-overview}
\newcommand{\featureexplanation}{https://visual-musicology.com/corpus/?usecase=feature-explanation}

\newcommand{\hllink}[1]{\textcolor{blue}{\underline{#1}}}
\newcommand{\usecaseA}{\textbf{\href{\epochcomparison}{\texttt{[UC1]}}}}
\newcommand{\usecaseB}{\textbf{\href{\atonality}{\texttt{[UC2]}}}}
\newcommand{\usecaseC}{\textbf{\href{\composercomparison}{\texttt{[UC3]}}}}

CorpusVis offers multiple prepared analysis use cases consisting of preselected data and specific component configurations tailored to the analysis of the respective topic. Users can create additional scenarios by either manually filtering compositions from the prepared dataset compositions or uploading their own datasets. In the remainder of this section, we will introduce three exemplary use cases. The first two use cases represent a hypothetical user who is interacting with the system while \usecaseC~has been provided by one of our expert users, showcasing how CorpusVis covers the introduced tasks. All use cases can be opened directly in CorpusVis via the following links: \href{\epochcomparison}{\hllink{Investigating Epoch Characteristics}}~\usecaseA, \href{\atonality}{\hllink{Tonality vs. Atonality}} \usecaseB, and \href{\composercomparison}{\hllink{Composer Comparison}}~\usecaseC.

\subhead{\href{\epochcomparison}{Investigation of Epoch Characteristics~\usecaseA}  --} 
In this use case, the analyst wants to explore the differences between the historical epochs~\taskB. 
Thus, from the list of use cases, the analyst selects \emph{Epoch Comparison}. 
This use case considers 51 heterogeneous compositions from composers like \emph{Pachelbel} (baroque), \emph{Wagner} (romantic), and \emph{Schönberg} (modern). Here, only a subset of eight features of the three feature categories (M, P, R)~\bLabel~such as \emph{Repeated Notes} and \emph{Pitch Variety} are selected that separate the four epochs within the projection view~\taskC. While the modern and baroque compositions are distinguishable from the all others, the pieces from the intermediate epochs classic and romantic have more overlap~\dLabel. The analyst identifies composers between the romantic and modern epoch, such as \emph{Stravinsky}, \emph{Rachmaninoff}, \emph{Schönberg}, and \emph{Webern}, that reside within classical and romantic composer groups. By using the column of the feature \emph{Size of Melodic Arcs} to sort the feature matrix, the analyst detects that romantic composers (yellow circles in the projection view) often used large melodic arcs compared to artists from other epochs. 

\subhead{\href{\atonality}{Atonality versus Tonality~\usecaseB} --} 
An interesting aspect regarding the harmonic development between the different epochs is the discussion of tonality and atonality. Atonal music is lacking a tonal center which means that none of the twelve available pitch classes is favored over the others as it is typically the case for classical music. Moreover, atonal music comprises more dissonant intervals (such as minor second and tritone intervals) which are available in the pitch feature group~\bLabel. Specifically low values of the feature \emph{Most Common Pitch Class Prevalence} are a good indicator for a composition to be atonal. In this example, the analyst is presented with 18 compositions from both classical composers including Mozart and Beethoven who composed rather tonal music as well as composers like Schönberg and Webern who are well-known by musicologists for their atonal compositions~\aLabel. After starting with the analysis, the analyst uses the projection view~\dLabel~to see which compositions are positioned closer together to detect similar pieces~\taskE~(see~\autoref{fig:projection_view}). As described, the analyst knows that tonality depends more on pitch and melodic features and not on rhythm. The configuration of the feature matrix~\bLabel~in this use case does not contain any rhythmic feature, which covers the mental model of the analyst. By viewing the columns in the feature matrix, the analyst detects as expected that the \emph{Melodic Tritones} feature has much higher values for the atonal compositions confirming the hypothesis that atonal compositions contain more dissonant intervals~\taskB. By investigating the feature distribution chart of \emph{Number of Common Pitches}, the analyst discovers that the compositions subset of the configuration comprises more pieces with zero common pitches compared to the overall distribution in the corpus~\taskC. Since the correlation matrix confirms a positive correlation of the \emph{Number of Common Pitches} to \emph{Most Common Pitch Class Prevalence}, the analyst is interested to see the values for the single compositions. The analyst is able to detect that the pieces by Schönberg and Webern try to avoid a tonal center as none of the available pitches seems to be favored compared to the works by the classical composers~\taskE.

\subhead{\href{\atonality}{Composer Comparison~\usecaseC} --} 
In the first scenario, the domain expert compares similar pieces from two different composers~\taskC and selects \textit{Erlkönig}\taskA by Franz Liszt and Franz Schubert~\aLabel. 
As a first step, the analyst manually inspects the first page using the sheet view~\eLabel~and detects the use of melodic octaves in the left hand of Liszt's interpretation which appears to be a salient difference between the compositions.
Based on this observation, the expert focuses on the features \emph{Melodic Octaves} and \emph{Average Melodic Interval} which he assumes that these features should reflect this characteristic. 
The feature matrix \bLabel~confirms the analyst's initial finding as the values for the \emph{Melodic Octaves} feature is 28.1\% for Liszt, but only 2.05\% for Schubert~\taskC.
Consequently, the analyst wants to explore whether this aspect is a typical difference between the two composers~\taskB and selects all available pieces in the dataset from both composers and aggregates them based on the composer attribute~\taskD.
Although the distinction decreases, the trend is still present (6.30\% - Liszt vs. 3.8\% - Schubert).
While the two features selected by the user are correlated, the analyst discovers that they represent discriminating characteristics between Liszt and Schubert. 
As the last step, the analyst is eager to see any unusual pieces (outliers) and ungroups the two clusters back into the single compositions.
By activating the clustering in the MDS projection view~\dLabel~the analyst groups similar pieces into groups finding many similarities on the feature level between Liszt and Schubert~\taskE.
For instance, while the compositions \textit{11. Ecossaise D.781} (Schubert) and \textit{Vive Henri IV} (Liszt) are quite similar on the feature level. The user finds \textit{Ave Maria - Ellens dritter Gesang} by Schubert to be an interesting outlier compared to most of the other compositions.
Eventually, the analyst uses the metadata table~\cLabel~to sort the matrix using the composer attribute to explore differences between the two collections~\taskB.
The analyst confirms the less surprising hypothesis that Liszt's works have a much higher \textit{Pitch Variety}.
Based on the feature \emph{Number of Common Pitches} he discovers that while Liszt almost always has zero pitches which individually account for at least 9\% of all notes, Schubert even has compositions with five common pitches such as \emph{Nachtviolen (D.752)}~\taskC.

\newcommand{\smeA}{\p{SME}{1}} 
\newcommand{\smeB}{\p{SME}{2}} 
\newcommand{\smeC}{\p{SME}{3}} 
\newcommand{\smeD}{\p{SME}{4}} 
\newcommand{\smeE}{\p{SME}{5}} 
\newcommand{\studA}{\p{ST}{1}} 
\newcommand{\studB}{\p{ST}{2}} 
\newcommand{\studC}{\p{ST}{3}} 
\newcommand{\studD}{\p{ST}{4}}

\newcommand{\p}[2]{{$\mathbf{#1}_\mathbf{#2}$}}

\section{Pair Analytics Study}
We conducted pair analytics sessions with users of different expertise levels to evaluate the applicability of \corpusvis~introduced in~\autoref{sec:workspace}. Since the number of available qualified domain experts is limited, we decided to use the pair analytics study method~\cite{pairanalytics2011}.

\subhead{Preliminary Expert Feedback --}
Before the actual study, we presented \corpusvis~to three domain experts to gather initial feedback in an informal pre-study.
The domain experts appreciated interacting with an exploratory  visualization. One of the experts liked the way on how music information could be retrieved through simply entering keywords or mouse interactions on the composer timeline or types~\taskA.
The interlinked components helped in understanding the connection between low-level features and the compositions~\taskB.
They found it useful to explore similarities and differences between composers and types by grouping them together~\taskC or through using the clustering in the projection view~\taskD. They appreciated the multiple components and the use of color to visually show connections of similar entities~\taskE. Especially, they could use the feature filtering functionality to identify outliers within single feature dimensions. They also noted that the detailed description of the features in connection with the sheet view provides music students the opportunity to experience new characteristics of sheet music~\taskB. 
In their opinion, the sheet music component is a core element to verify hypotheses and to investigate the quality of compositions. However, they also critiqued our used color map and the lack of a feature to upload their own music sheets. Both of these issues were addressed in a refinement cycle before our study. In addition, based on our observations from the interactions in the pre-study, we added the features of interactive clustering and distribution-based grouping and filtering.  

\subsection{Study Design}
We conducted the study to gather user feedback to identify benefits and drawbacks of the introduced analysis workspace. For the evaluation, we used a video conferencing software to show the interactive workspace to the participants during the study trials which enabled them to easily interact with~\corpusvis~ensuring equal study conditions. Based on the elicited \emph{task characteristics} described in~\autoref{sec:problemdescription}, the participants carried out different analysis questions provided by the study director. From the study results, we gained useful insights on our work and its potential impact.

\subhead{Study Participants --}
In sum, nine participants with different expertise levels partook in the study. Due to the required musical knowledge to use the analysis application, we focused our study on the target user group of domain experts including musicologists and music teachers. Five participants had an experience level of at least ten years including four university professors and a music teacher, which we further refer to as \textbf{s}ubject \textbf{m}atter \textbf{e}xperts (\smeA--\smeE). The remaining four participants were Ph.D. students (\studA--\studD) who are professionally involved with music on a daily basis but with less experience about music analysis compared to the SMEs. 

\subhead{Methodology --}
We applied a combined \emph{observation} and \emph{pair analytics} study~\cite{pairanalytics2011} to investigate the usefulness of \corpusvis. In total, we performed nine study trials (average duration of 95 minutes) that were each led by a visual analytics expert (VAE) of our team. Each study session was structured into four sections. \\
The first part started with an \emph{introduction} into the topic of sheet music collection analysis to frame the context of our work. Each participant completed a \emph{demographic questionnaire} to provide information about their age, gender, level of education, and their level of expertise regarding music analysis. The first phase ended with an interview about the participants' \emph{expectations} regarding an analysis system intended to support the investigation of sheet music collections. 
\begin{figure}[t]
    \centering
    \includegraphics[width=\linewidth]{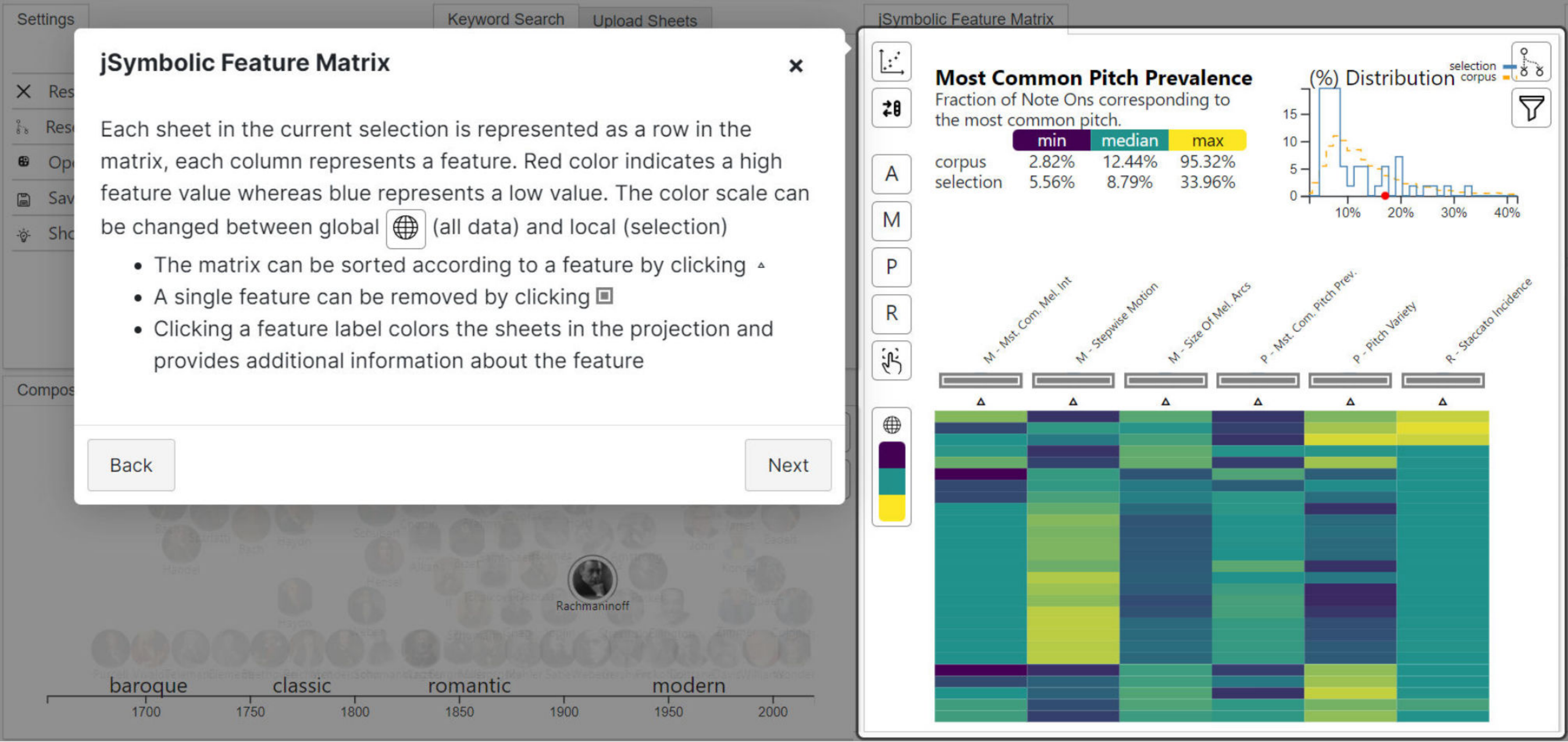}
    \xtramargin{\vspace*{-15pt}}
    \caption{
    We used Intro.js~\cite{introjs2016mehrabani} to prepare a visual explanation of all components of our prototype to ensure a consistent introduction of the interactive analysis workspace for all study participants.
    }
    \label{fig:introjs}
    \xtramargin{\vspace*{-2em}}
\end{figure}
After the general introduction, each participant received a detailed, textually guided \emph{walkthrough} regarding all components of the study prototype (see~\autoref{fig:introjs}). Then, we introduced the five analysis tasks \taskA--\taskE from the analyzed requirements to each participant. Besides the analysis tasks, the introduction comprised an explanation of the relevant metadata features, including composer, type, epoch, and the low-level features extracted from the musical compositions visualized by the different components. In addition, the participants were informed about the origin of the used dataset (MuseScore~\cite{musescorecolunteered2016}) and its content addressing the overall amount of considered composers and musical pieces. Phase 2 closed with a simple and unguided exploration (similar to~\taskB) of the visual interactive workspace, allowing the participants to get acquainted with the application. This introductory task contained a preselected set of ten different compositions and five features. As a quick entry point, we prepared a use case \emph{\href{\featureexplanation}{feature explanation}} that can be directly loaded from the use case selector in the interface the provides this preselection.
We chose the pieces based on contrasting (extreme) feature values such as low and high \emph{Most Common Pitch Prevalence} values (3\% and 65\%). The participants were asked to explore the \emph{Feature Matrix} (see~\autoref{fig:jsymbolicFeatureMatrix}) to sort the selected sheets according to features of their interest. Then they used the \emph{sheet view} (see~\autoref{fig:sheetview}) to relate the feature values to the underlying common music notation. \\
During the main part of the study, we asked the participants to execute the analysis tasks, as described in the next section.
We encouraged them to verbalize their thoughts and interactions (\emph{think-aloud concept}). After completing the prescribed analysis tasks, we performed an interview to receive feedback about the participants' experience regarding the functionality and usability of the prototype. We closed the study trial with a final questionnaire containing nine questions answered via a five-point Likert scale to identify the benefits and drawbacks of \corpusvis. 

\subhead{Study Tasks --} During the pair analytics session, all participants analyzed sheet music collections based on three use cases introduced in~\autoref{sec:workflows}. \usecaseA~comprised 51 compositions from various composers of all epochs. After getting acquainted with the metadata of the selected pieces, the participants selected all pieces from a single epoch of choice~\taskA~and identified at least two discriminating features~\taskC. For this task each participant used the feature matrix for identifying similarities between the compositions~\taskE. In addition, the participants could add additional pieces through the keyword search to extend the selection~\taskA. We encouraged each participant to freely explore the visualizations with individual compositions of interest~\taskB.\\
A second task~\usecaseB addressed the comparison of \emph{atonal} with \emph{tonal} pieces (Tonal versus Atonal), which used contemporary and classical works. Similarly, the participants were asked to detect characteristic features values for both groups by comparing the music sheets in the projection view and the feature matrix~\taskC. The last task addressed comparing two or more composers relevant to the participant. For this, the \emph{Composer Timeline} helped filtering the respective composers~\taskA, followed by a grouping of all pieces of single composers~\taskD. Then, we asked the participant to visually examine the feature differences~\taskE of the composers~\taskB.

\subsection{Evaluation Results}
We gathered feedback from the qualitative evaluation during the different study parts that we grouped thematically in the following.

\subhead{Participants' Expectations --}
During the expectation interview, the participants stated various aspects of an application intended to support the musical analysis of sheet music corpora. \studA~remarked that he expected the application to group a musical repertoire into genres and composers automatically. \smeA~expected the analysis system to use all information available in musical scores to perform musical analysis, which should present differences between musical styles, since ``\textit{classical is different from pop music}''.\smeB~mentioned that she expects an interface that allows her to provide information about her interests. Then, the ``\textit{system would automatically provide analysis results without the need for manual analysis}''.

\subhead{Exploratory Analysis --}
To familiarize themselves with the compositions of \usecaseA, the participants started that by investigating the metadata table to see which composers are selected. Applying the color scheme of the composer timeline to the projection view revealed the  similarity of the compositions. \smeA~mentioned the separate grouping of contemporary Jazz and Pop pieces compared to Baroque and Classical compositions, which also were separate groups but had some overlapping. Using the lasso to select the yellow-colored (modern) pieces, she identified that compared to the rest, these pieces have more note repetitions and high proportion of consonant melodic intervals and less minor/major second intervals. \studD~checked the quality of a sheet file based on features such as pitch variety.  Typically complex compositions such as etudes and symphonies have a high pitch variety. Thus, participants could delineate custom arrangements from more original compositions. \\
While analyzing the configuration of \usecaseB, the participants were presented with tonal compositions from composers such as Mozart and Beethoven, which were compared with atonal pieces from Schönberg and Webern~(see~\autoref{fig:projection_view}). The participants used the projection view to compare similarities and appreciated the sorting functionality in the feature matrix, which helped them confirm that atonal pieces use tritone intervals more often. They identified that the feature \emph{Number of Common Pitches} for the atonal compositions is always zero, which reflects that composers objective not to favor certain pitch classes. Especially \smeB~appreciated the sheet view, which allowed her to compare the values from the feature matrix to the common music notation. The third use case started with aggregating all compositions from the composers displayed in the composer timeline. The participants found that \emph{Changes Of Meter} is the most discriminating feature revealing that contemporary music seldomly changes the meter signature within a composition. 

\subhead{Comparative Analysis --}
The participants used the feature matrix to identify similarities between single compositions or composer groups. 
\smeD~and \smeE~found the melodic features useful to investigate the melodic content. In addition, \smeC~and \studA~appreciated the features regarding the intervals, which they considered to be particularly helpful to identify dominant melodic aspects. \smeA~mentioned that the matrix helped confirm that contemporary music such as ``Imagine Dragons and Coldplay is more consonant and has more note repetitions, but fewer stepwise motion'' compared to Schönberg (Op. 11) or Webern (Op. 4). Using \corpusvis, \smeA~learned that minuets from Bach have much lower \emph{Pitch Class Variety} compared to his fugues. \studA~noted that based on the available pieces, Schönberg's work has more chromatic motion (25\%) compared to Webern (20\%). \studB~and \studC~emphasized that the pitch range and density information helped them investigate the pitch material to better understand the complexity without even considering the sheet view. \studC~could identify that Beethoven uses 20 fewer pitches compared to Liszt, who, as he stated, uses more chromatic melodic progressions. During the investigation, the participants found different matrix features to be relevant. They could easily understand the separation into the three categories rhythm, pitch, and melody, which were of different importance depending on the task. For instance, the time signature is a salient feature that helps to determine certain composition types. The pitch variety helped them detect complex compositions and differentiate tonal from atonal compositions. The chromatic motion feature supported identifying twelve-tone pieces.

\begin{figure}[b]
    \xtramargin{\vspace*{-10pt}}
    \centering
    \includegraphics[width=\linewidth]{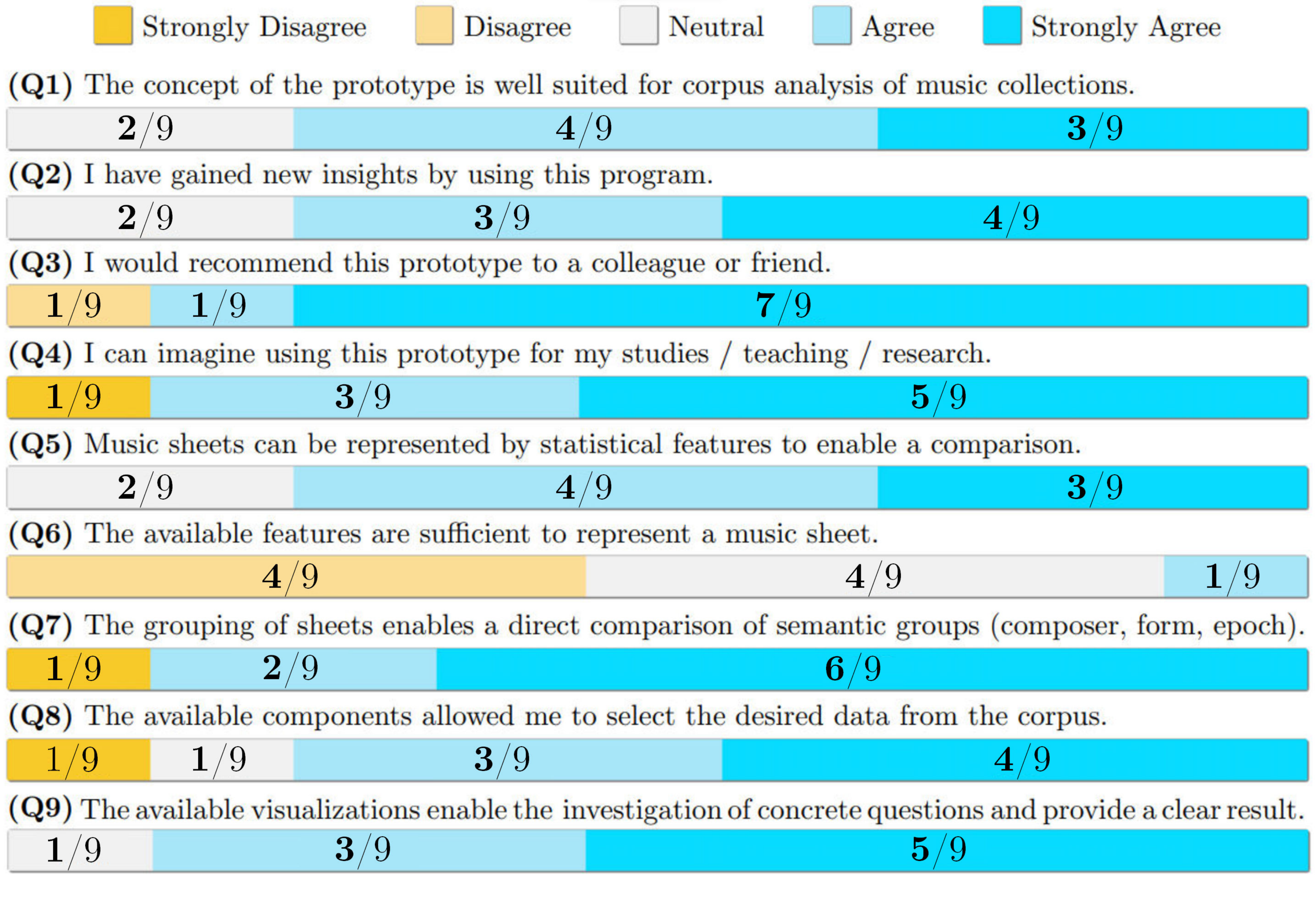}
    \xtramargin{\vspace*{-15pt}}
    \caption{
    The final questionnaire comprised nine questions to assess the usability and functionality of \corpusvis~for analyzing sheet music collections. Most questions achieved good ratings, emphasizing an overall high satisfaction level.
    }
    \label{fig:questionnaire}
    \xtramargin{\vspace*{-15pt}}
\end{figure}

\subhead{Closing Questionnaire --}
All participants completed a final questionnaire answering nine questions to assess the usability of \corpusvis. 
The questions and the results that are based on a 5-point Likert scale ranging from ``Strongly Disagree'' to ``Strongly Agree'' are displayed in \autoref{fig:questionnaire}. The questionnaire results reveal that our participants gained new insights (Q2) about sheet music collections by enabling them to perform groupings of the dataset to carry out comparison tasks based on epoch, composers, and composition type information~(Q7). Most participants agreed that \corpusvis~is suitable for the visual analysis of sheet music corpora~(Q1, Q9). Q6 received the lowest scores, which illustrates that the computationally extracted features do not sufficiently address all users' aspects, such as dynamics, instrumentation, or harmonic progressions. Nevertheless, the scores of Q5 show that the computed statistical features still enable the comparison even though they may contain all aspects the analysts would be interested in. All but one of the participants stated that they would recommend the interactive workspace to a colleague or use it for their studies and research~(Q4). For most participants, the currently used dataset in \corpusvis~contained the compositions they wanted to add to their analysis~(Q8). Overall, the questionnaire results convey that the participants found potential for analyzing sheet music collections and would continue using it in the future, especially if it is improved based on this study's outcomes.

\section{Discussion}
\label{sec:discussion}
We identified benefits and drawbacks from the study which showed that \corpusvis~supports specific analysis tasks such as comparing composition types or even the entire work of composers. Many participants questioned the quality of the underlying MuseScore dataset which contains volunteered sheet music. To address this issue, \corpusvis~enables the upload of sheet music collections, allowing musicologists to use their own, curated datasets if desired. In this way, we mitigate the negative influence of datasets that suffer from poor quality. Despite the underlying dataset's limited quality, \corpusvis~demonstrates the positive impact that tailored visualizations have on the work of musicologists.

Musicologists controversially discuss whether and where to place clear borders between different epochs of music history. In our approach, we simplify the timeline of musical history into four major epochs \emph{baroque}, \emph{classic}, \emph{romantic}, and \emph{modern}. While we assign individual colors to each epoch, study participants appreciated that we introduce smooth color gradients to emphasize the transitions period. During the different musical periods, the focus on specific characteristics of music has shifted. While \corpusvis~already includes several features, there may be other music styles that would require to extend the application. For instance, starting from the second half in the 20$^{\mathrm{th}}$ century, new musical frontiers emerged, including serial, electronic, minimal, spectral, and experimental music. These contemporary music styles sometimes may not be available through transcribed common western music notation. They often rely on auditory or physical features such as timbre or spectral data. Therefore, such music compositions may only be available through audio files but not symbolic music. \corpusvis~would either require an audio extension or a suitable symbolic notation that can be used to extract features that represent the underlying information.

During the implementation and design of the presented interactive workspace, we have encountered several challenges. A typical issue in this field is \emph{copyright}. As MuseScore users sometimes upload compositions from contemporary musicians or artists who are still alive, these pieces may be deleted from MuseScore again. To avoid legal issues caused by sheet music that should not be visible in our system, we only include compositions from artists who died more than 70 years ago that are still available on MuseScore. Whenever sheets are deleted from MuseScore for any reason they are also no longer available in our system.

During the development of this system we were in close contact with multiple domain experts from musicology, allowing us to gather early feedback on~\corpusvis. We quickly identified challenges in this inter-disciplinary collaboration, especially in creating visualizations that were understandable by experts while allowing them to focus on aspects relevant to their intended analysis. In particular, we found that the field of music analysis is extremely broad, necessitating a flexible system with a myriad of different options. However, adding additional functionality also complicated the system's design, in turn making it more difficult for experts to understand and use. As a result, in this work we focus on specific tasks and data characteristics to limit the scope.

\vspace{-2pt}
\subsection{Limitations}
While the study results show that \corpusvis~supports different music analysis tasks, it also surfaced some limitations.
For instance, some composers of the 20$^{\mathrm{th}}$ or 21$^{\mathrm{st}}$ century heavily use timbral or textural elements. These features are often not encoded using the common Western music notation at all. As a result, \corpusvis~cannot be of help in analyzing work from such contemporary composers. Future extensions of \corpusvis~could integrate auditory features derived from recordings rather than only music sheets to capture the expression with which musicians played the compositions to close this gap. In its current implementation, \corpusvis~does not capture the instrumentation of a composition. Several participants (\smeB--\smeE) stated that they would like to filter all compositions of a composer that contain piano or stringed instrument parts. Especially, identifying how the pitch range is distributed between different instruments would provide useful semantic information about the role of instruments and overall timbre of a piece. 

While \corpusvis~contains works from several composers, the visualization only partly scale based on the available aggregations. 
The projection view and feature matrix scale towards adding additional compositions from the same composer using aggregation based on the composers as it uses the equal visual space. 
Yet, an analyst may need to focus the analysis on a subset of musical features if other characteristics should be considered as the cognitive complexity increases with the number of features to be analyzed. 
Nevertheless, \corpusvis~already uses semantic groups for melodic, rhythmic, and pitch features to enable users to focus on a subset.
To improve this, \corpusvis~could be adapted to allow users to create custom feature groups to focus the analysis and additional features that could be easily added if required. 
The application's performance depends on the library music21, which loads every composition once at the beginning to prepare all features for visualization.

\smeB~mentioned that the classification of the pieces into their types may be ambiguous, and that enabling analysts to re-classify compositions could increase the analyst's influence on a dataset's quality besides the option of uploading custom datasets. \studA~and \smeA~remarked that the vertical combination of notes (simultaneously played notes is referred to as harmony) is not covered by the available features, since the employed jSymbolic features are limited to pitch, melodic, and rhythm characteristics. Although \corpusvis~already offers a large array of selection and filtering mechanisms, some participants would have liked additional options for filtering a dataset by key signatures, voicing, and lyrics.

\subsection{Take-Home Messages}
From our work, we distill three primary take-home messages.

\textbf{(1) Corpus-level music analysis is challenging due to the magnitude and complexity of underlying features.} 
Manual analysis of sheet music collections is a tedious process requiring profound domain knowledge. 
\corpusvis~supports the analysis of extensive sheet music collections, which is not feasible when done manually. 
While abstract visualizations open up new perspectives, they should not be considered a replacement for traditional analysis methods but rather an extension. 
Thus, integrating domain experts during the design phase and the requirements analysis is crucial to address their specific needs.
This enables users to combine close- (sheet view) and distant-reading (projection, features) and get a corpus-overview.

\textbf{(2) Visualization dashboards can enable multiple perspectives on the data, supporting scalable and multi-faceted analysis.} 
Especially when musicologists want to explore, understand, and interpret music sheet corpora, visual analytics helps addressing the individual needs of analysts.
Automatic computational methods do not provide visualizations to users facilitating interactive engagement with a sheet music collection to understand its underlying characteristics better. 
We allow for various interconnected analysis workflows by tailoring the components' visual design to their targeted task.
The availability of multiple components covering different characteristics of the dataset enables users to set their analysis focus on those attributes they are interested in their analysis.

\textbf{(3) Tailored interactions enable a task-driven analysis of corpora using diverse metadata and statistical features.}
Interactive visualization allows investigating sheet music from different perspectives, supporting various aspects music analysts may be interested in. In particular, for analyzing complex questions involving multiple meta-data and features, we rely on the interactive inter-linking of visual components, enabling an expressive yet visually uncomplicated analytics system.   
Enabling analysts to import their own curated datasets facilitates their investigation and generation and confirmation of hypotheses.

\subsection{Future Work and Research Opportunities}
We postulate that our work only represents a starting point for future cooperation between visualization and musicology researchers, and we want to emphasize that it is worth ``risking the drift'' for such interdisciplinary collaborations~\cite{hinrichs_risk_2017}.
So far, the classification of the compositions metadata available in \corpusvis~is fixed and cannot be adjusted by the analyst. 
In the future, it may be worthwhile to enable users to change meta information to increase the dataset quality. Extending the application through labeling would further improve the analysts' ability to capture and externalize analysis results. 

While we have considered the temporal aspects of music, we have neglected the geographic information of where composers have lived and worked which has been already studied by Khulusi et al.~\cite{musixplora2020khulusi}.
It would be interesting to investigate how the location influenced the style of specific composers.
While the low-level characteristics are just statistical measures, it would be interesting to see how they correlate with the mood and emotions of listeners. So far, we have not considered the dynamics, instrumentation, or timbre of the instruments that the compositions are written for, which would be worthwhile to add support for in the future.
If pianists or guitarists are looking for new pieces to play, it would be helpful to provide specific instrument filters that help them find relevant information. Moreover, as the dataset's quality plays a primary role in analysis, it would be helpful to provide classification means for analysts to identify low-quality pieces to exclude unwanted items. Eventually, \corpusvis~focuses the analysis on a high level. We recently published \href{https://visual-musicology.com/sheetmusicvis/}{{MusicVis}} that enables analysts to investigate single compositions at the sheet level~\cite{MiFuHa2021Augmenting}. We aim at connecting both approaches to allow analysts to filter relevant compositions at a higher level using~\corpusvis~which can then be further analyzed at the sheet level.

\section{Conclusion}
In this paper, we have introduced \corpusvis, a \href{\appurl}{web-based interactive workspace} for the visual analysis of sheet music collections.
We discussed data and task characteristics relevant to analyzing sheet music corpora.
Multiple connected components through linking and brushing support exploration, detection, and comparison tasks.
Based on initial expert feedback, we identified benefits and challenges of our approach.
In addition to three exemplary use cases that demonstrate the analysis workflow to compare composers and types, we conducted a pair of analytics study with nine participants. 
The qualitative study results show that music analysts benefit from interactive visualization addressing different use case scenarios such as comparing composers, epochs, and composition types based on low-level features including pitch, rhythm, and melody.

\clearpage

\bibliographystyle{eg-alpha-doi}
\bibliography{references/basic,references/specific,references/misc}
	
\end{document}